\documentclass[preprint,onecolumn,aps,showpacs,showkeys,nofootinbib,superscriptaddress]{revtex4}
\usepackage{amsmath,amsfonts,bbm}
\usepackage{graphicx}

\begin{document}
\title{NLO QCD corrections to $ZZ$+jet production at hadron colliders}
\date{\today}
\author{T.~Binoth}
\affiliation{School of Physics and Astronomy, University of Edinburgh, Edinburgh EH9 3JZ, UK}
\author{T.~Gleisberg}
\affiliation{SLAC National Accelerator Laboratory, Menlo Park, CA 94025, USA}
\author{S.~Karg}
\affiliation{Institut f\"{u}r Theoretische Physik E, RWTH Aachen, 52056 Aachen, Germany}
\author{N.~Kauer}
\affiliation{Department of Physics, Royal Holloway, University of London, Egham TW20 0EX, UK}
\affiliation{School of Physics and Astronomy, University of Southampton, Southampton SO17 1BJ, UK}
\author{G.~Sanguinetti}
\affiliation{LAPTH, Universite de Savoie, CNRS, BP. 110, 74941 Annecy-le-Vieux, France}
\begin{abstract}
A fully differential calculation of the next-to-leading order QCD corrections to the 
production of $Z$-boson pairs in association with a hard jet at the Tevatron and LHC 
is presented.  This process is an important background for Higgs particle and 
new physics searches at hadron colliders.  We find sizable corrections for cross sections 
and differential distributions, particularly at the LHC.  Residual scale uncertainties 
are typically at the 10\% level and can be further reduced by applying 
a veto against the emission of a second hard jet.
Our results confirm that NLO corrections do not simply rescale LO predictions.
\end{abstract}
\maketitle
\preprint{Edinburgh 2009/18, LAPTH-1362/09, PITHA 09/30, SFB/CPP-09-113, SLAC-PUB-13833}


\section{Introduction\label{intro-section}}

Weak boson pair production at hadron colliders plays an essential part
in the search for Higgs particles and for physics beyond the Standard Model 
(SM), since weak bosons can decay into jets, charged leptons or neutrinos 
and hence produce the same signatures as Higgs bosons, new coloured particles, 
new electroweak gauge bosons or dark matter candidates.  In addition to
being an important background to direct new physics searches at the 
Large Hadron Collider (LHC) \cite{Campbell:2006wx}, weak boson pair production 
also allows to search for new physics via experimental evidence for SM
deviations in the form of anomalous interactions between electroweak gauge bosons
\cite{anomalous_general}.

$Z$-boson pair production has been observed at the Tevatron \cite{Aaltonen:2008mv}
and studied extensively by the LHC general-purpose detector collaborations 
\cite{atlas_cms_tdrs}.
Since LO predictions for hadron collider processes are affected by large QCD scale 
uncertainties with respect to normalisation and kinematical dependence, the 
inclusion of NLO QCD corrections is important when comparing predictions for 
cross sections and differential distributions with data.
Theoretical predictions for $ZZ$ production 
at leading order (LO) \cite{Brown:1978mq} have thus been improved by including
next-to-leading order (NLO) QCD corrections without \cite{ZZ_NLO} and with decays \cite{ZZ_NLO_decays}.
More recently, $Z$-boson pair production at NLO has also been 
investigated in selected SM extensions \cite{ZZ_NLO_BSM}.
As a window to new physics $ZZ$ production is particularly interesting
because of the absence of $ZZ\gamma$ and $ZZZ$ couplings \cite{N3GB}
in the SM.  Probing such anomalous neutral gauge boson couplings at hadron 
colliders has also been studied in the literature
\cite{anomalous_ZZV_studies}.

Going beyond final states with two particles, NLO QCD corrections 
have been computed for the production of three weak/vector bosons 
\cite{VVV_NLO}, the production of a weak boson in association with 
up to three jets \cite{VplusJets_NLO} and weak boson pair production 
in vector boson fusion \cite{VV_VBF_NLO}.  Of particular interest
is also the production of weak boson pairs with one additional
jet at NLO.
This process is interesting in its own right, due to the enhanced 
jet activity, particularly at the LHC.  It also provides the 
real-virtual contribution
to the next-to-next-to-leading order (NNLO) corrections to weak 
boson pair production.  The production of $W$-boson pairs with
an additional jet has thus been calculated at NLO without
\cite{Dittmaier:2007th} and with 
\cite{Campbell:2007ev,Dittmaier:2009un} decays.
An additional contribution to the NNLO corrections that has been
calculated for $WW$ and $ZZ$ production (at LO) including decays
is the loop-induced gluon-fusion process \cite{ggVV}.
Other building blocks for the NNLO calculation of weak boson 
pair production have been presented in Ref.~\cite{VV_NNLO_other}.

In this paper, we present a calculation of the ${\cal O}(\alpha_s)$ 
corrections to $Z$-boson pair production with an additional jet at 
hadron colliders in the SM.\footnote{%
Partial results of our calculation have already 
been presented in Ref.~\cite{ZZj_NLO_partial}.
}
Details of the NLO calculation are described in Sec.~\ref{nlo-section}.
We then present numerical results in Sec.~\ref{results-section}
and end with conclusions.


\section{Details of the NLO calculation\label{nlo-section}}

At LO, all channels for $ZZj$ production at hadron colliders 
are related to the amplitude $0\to ZZq\bar{q}g$ by crossing
symmetry.  Therefore, the following subprocesses 
contribute:
\[ 
q\bar{q} \to ZZg\,,\quad
qg \to ZZq\,,\quad
\bar{q}g \to ZZ\bar{q}\,,
\]
where $q$ can be either an up- or down-type quark.\footnote{%
The down-type quark initiated amplitudes are obtained from the
up-type quark initiated amplitudes by adjusting the chiral
couplings.}
We calculate in the 5-flavour scheme, i.e.~$q=u,c,d,s,b$, and
neglect all quark masses.
\begin{figure}
\begin{center}
\includegraphics[height=3.cm,angle=0,clip=true]{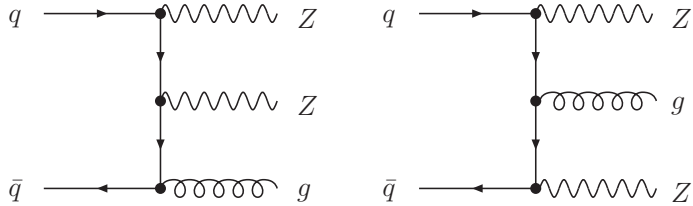}\\[0.2cm]
\caption{%
Representative LO graphs for the partonic process $q\bar{q}\to ZZg$.
\label{fig:lo-graphs}}
\end{center}
\end{figure}
Representative LO diagrams for the first subprocess are shown 
in Fig.~\ref{fig:lo-graphs}.  Comparing the $ZZ$ with
the corresponding $WW$ production amplitude, key differences are
that the $W$-boson coupling to fermions is purely 
left-handed, which leads to a reduced number of helicity amplitudes
in that case, the produced weak bosons are distinct for $WW$ production,
but identical for $ZZ$ production (leading to a significant 
increase in the number of Feynman diagrams), and graphs with 
triple-gauge boson vertices contribute to $WW$,
but not to $ZZ$ production.

\subsection{Virtual corrections\label{virt-subsection}}

\begin{figure}
\begin{center}
\includegraphics[height=6.5cm,angle=0,clip=true]{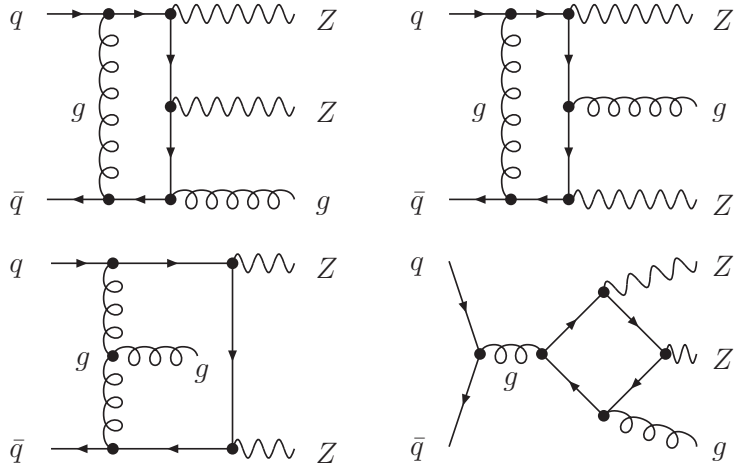}\\[0.2cm]
\caption{%
Representative one-loop graphs for the partonic process $q\bar{q}\to ZZg$.
\label{fig:loop-graphs}}
\end{center}
\end{figure}

At ${\cal O}(\alpha_s)$, the most complicated loop topologies 
are pentagon graphs derived from the tree-level
graphs via virtual gluon exchange (and crossing)
and box graphs derived by closing the quark line in the 
tree-level graphs and attaching a $gq\bar{q}$ current.
Since we calculate
with $N_f=5$ and neglect quark mass effects, graphs with 
Higgs boson exchange do not contribute.
Triangle graphs, where three gauge bosons 
couple to a quark loop, also vanish in the
massless quark limit.
Representative one-loop graphs for the partonic process $q\bar{q}\to ZZg$
are shown in Fig.~\ref{fig:loop-graphs}.
Starting from the Feynman graph representation, two independent sets of 
amplitude expressions have been generated:~one manually, the other 
using QGRAF \cite{Nogueira:1991ex}.
Both representations employ the spinor helicity formalism of Ref.~\cite{Xu:1986xb}.  
Polarisation vectors have been represented via spinor traces, i.e.~kinematic
invariants up to global phases.  By obtaining an analytical representation for 
the full amplitude, we aim at promoting simplification via analytical cancellations.
Especially we employ that, apart from the rank one case, all pentagon tensor integrals 
are reducible, i.e.~can directly be written as simple combinations of box tensor integrals.   
For the remaining tensor integrals we employ the GOLEM-approach \cite{golem_reduction}.
In this approach, the use of 6-dimensional IR finite box functions
allows to isolate IR divergences in 3-point functions. 
We use FORM \cite{Vermaseren:2000nd} and Maple to obtain tractable analytical expressions for the 
coefficients to the employed set of basis functions for each independent helicity amplitude,
and to further simplify them.  The basis functions are evaluated using the GOLEM95 
implementation \cite{Binoth:2008uq}.
We note that for the reduction of box topologies one obtains the same result
as with the Passarino-Veltman tensor reduction \cite{passarino_veltman}. 
If one fully reduces all tensor integrals to a scalar integral representation, 
the difference between the two approaches results from the treatment of the pentagon 
integrals and the use of finite 6-dimensional box functions.

To treat $\gamma_5$ we employ the 't~Hooft-Veltman scheme 
\cite{Hooft_Veltman}, where the $\gamma^\mu$ are 
split into a 4-dimensional part that anti-commutes with $\gamma_5$ 
and a commuting remainder.\footnote{%
Note that the 't~Hooft-Veltman scheme treats the observed particles in 4 dimensions, but the
soft/collinear gluons in $d$ dimensions. This guarantees that for the IR subtractions
the same Catani--Seymour dipole terms as for conventional dimensional regularisation
can be used \cite{Catani:1996pk}.}
As is well known, to take into account differences between the QCD corrections 
to axial vector and vector currents, a finite renormalisation has to be performed.
To enforce the correct chiral structure of the amplitudes, 
a finite counterterm for the axial part is included in the used gauge boson vertex 
(see e.g.~Ref.~\cite{axial_anomaly_dimreg}):
\[
V^\mu_{Vq\bar{q}} \sim g_v \, \gamma^\mu + Z_5 \, g_a \, \gamma^\mu \gamma_5 \quad\text{with}\quad
 Z_5 = 1 - C_F \,\frac{\alpha_s}{\pi}\,.
\]

We have verified that the relative contribution of graphs with quark loops to  
the integrated results is well below 1\%.  We therefore neglect this contribution 
in the following.  We have compared our two independent implementations of the
virtual amplitudes -- both generated using the GOLEM reduction -- and have 
found agreement of 9-16 significant digits for all contributions 
at two test phase space points.  The discrete Bose, $P$ and $C$ symmetries 
induce relations between different helicity amplitudes that have been verified 
numerically as additional check.  We have furthermore tested gauge invariance 
by confirming the Ward identity for external gluons.  
Furthermore, we used the same tools as in the calculation presented here 
to calculate numerical results for the successful comparison of virtual 
corrections to $WWj$ production in Ref.~\cite{Bern:2008ef}.  That 
comparison therefore provides an additional check of our calculation.
We have also verified that our LO amplitude implementation is correct.
To calculate numerical results for the virtual contributions 
we employed the OmniComp-Dvegas package \cite{OmniComp}, which 
facilitates parallelised adaptive Monte Carlo integration
and was developed in the context of Ref.~\cite{OmniComp_references}.


\subsection{Real corrections\label{real-subsection}}

The ${\cal O}(\alpha_s)$ real corrections channels for 
$ZZj$ production at hadron colliders are related to the 
amplitudes $0\to ZZq\bar{q}gg$ and $0\to ZZq\bar{q}q'\bar{q}'$ 
by crossing symmetry.  While all virtual corrections channels 
are already active at LO, new real corrections channels 
open up at NLO, namely the $gg$, $qq'$, $q\bar{q}'$
($q'\neq q$) and $\bar{q}\bar{q}'$ channels.
Note that these new channels are effectively of LO type. 

To facilitate the cancellation of soft and collinear singularities
we employ the Catani-Seymour dipole subtraction method \cite{Catani:1996vz}.
We use the SHERPA implementation \cite{Sherpa} to calculate
numerical results for the finite real corrections contribution.  All
amplitude and dipole contributions have been verified 
through comparison with results calculated with two independent 
implementations:\footnote{%
Other implementations that automate the dipole subtraction method 
of Ref.~\cite{Catani:1996vz} have been reported in Ref.~\cite{other_dipole_implementations}.
}
MadDipole/MadGraph \cite{MadDipole_MadGraph}
and HELAC \cite{Helac}. In both comparisons 9-significant-digits
agreement or better was achieved for all contributions for two test phase space
points.  We have also successfully compared with an in-house 
implementation of a set of independent dipoles.


\section{Numerical results\label{results-section}}

In this section, we present first NLO predictions for 
$ZZj$ cross sections and differential distributions 
at the Tevatron ($p\bar{p}$, $\sqrt{s} = 1.96$ TeV)
and LHC ($pp$, $\sqrt{s} = 14$ TeV) and compare them with
the NLO results for $WWj$ production given in 
Ref.~\cite{Dittmaier:2009un}.\footnote{%
Differences in the input parameters are minute (less than 0.1\%).}

As mentioned above, our calculation employs the 5-flavour scheme
and the massless quark approximation (including the $b$ quark).
We use the SM parameters\footnote{%
We provide the exact input values of our calculation in order 
to facilitate reproducibility.
}
\[ M_Z = 91.188 \text{ GeV}\,,\quad \alpha(M_Z) = 0.00755391226\,,\quad \sin\theta_W^2=0.222247\,,
\]
and employ CTEQ6 parton distribution function sets \cite{Pumplin:2002vw}. 
LO (NLO) cross sections are calculated
with CTEQ6L1 (CTEQ6M) and LO (NLO) $\alpha_s$ running.  
For $\alpha_s(M_Z)$, the default
LHAPDF \cite{Whalley:2005nh} values are used: $\alpha_s(91.188 \text{ GeV}) = 
0.129783$ at LO and $\alpha_s(91.70 \text{ GeV}) = 0.1179$ at NLO.
In our parton-level calculation, partons are clustered into jets using the 
inclusive $k_t$ algorithm \cite{kt_algo} with $R = 0.7$.  
To study the scale dependence of cross section predictions
we use the discrete grid $\mu = 2^{i/2}M_Z$ with $i\in\{-7,-6,\dots,7\}$.  
The renormalisation and factorisation scales are identified here 
($\mu = \mu_R = \mu_F$).  We apply a $p_T > 50$ GeV
cut on the hardest jet unless noted otherwise.

\begin{figure}
\begin{center}
\begin{minipage}[c]{.49\linewidth}
\flushleft \includegraphics[height=9.5cm,angle=0,clip=true]{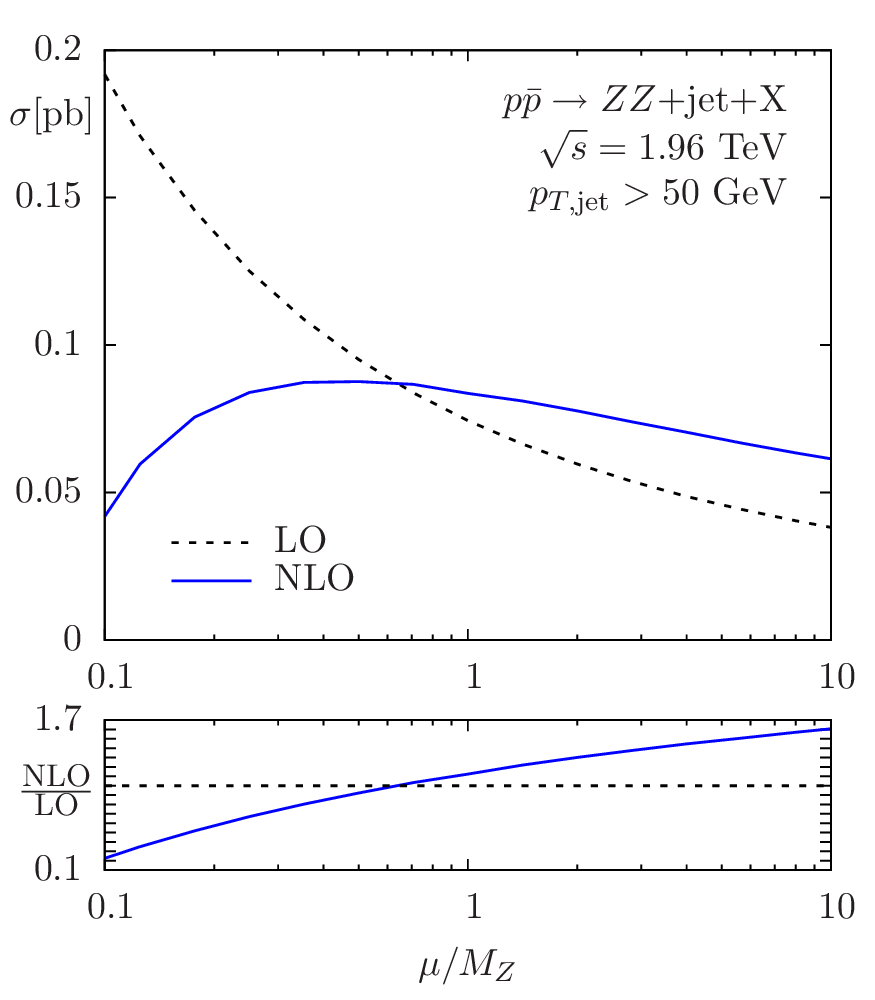}
\end{minipage} \hfill
\begin{minipage}[c]{.49\linewidth}
\flushright \includegraphics[height=9.5cm,angle=0,clip=true]{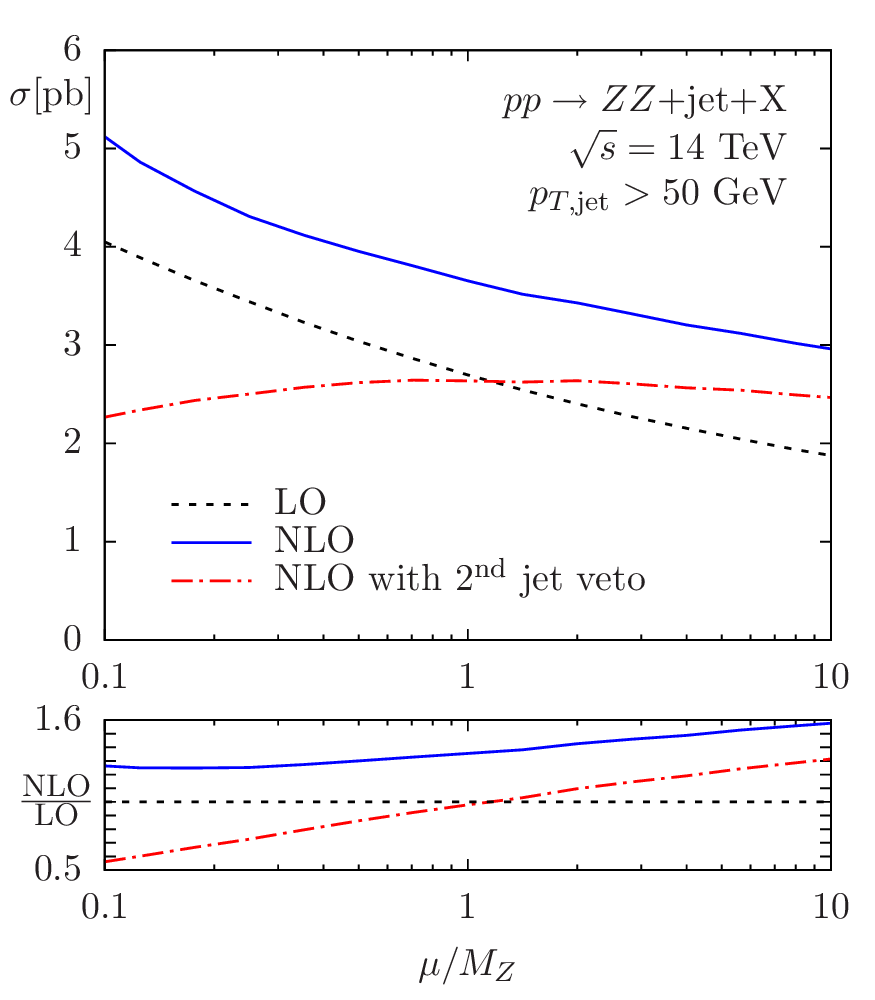}
\end{minipage}\\[0.2cm]
\caption{%
Comparison of the scale dependence ($\mu_R=\mu_F=\mu$) of the $ZZ$+jet cross section 
at the Tevatron and LHC with $p_{T,\,\text{jet}} > 50$ GeV for the hardest jet 
in LO (dotted) and NLO (solid).  For the LHC the exclusive NLO cross section when a
$p_{T,\text{jet}} > 50$ GeV veto for additional jets is applied is also
shown (dot-dashed).   Input parameters are defined in the main text.
\label{fig:scalevar}}
\end{center}
\end{figure}

In Fig.~\ref{fig:scalevar}, LO and NLO predictions for $ZZj$ production 
cross sections at the Tevatron and LHC are displayed as function of the
QCD scale $\mu$, which is varied by a factor 10 around the $Z$ mass.
At LO, we find a much larger scale variation at the Tevatron than at the
LHC.  When NLO corrections are included, the Tevatron cross section 
reaches its maximum at approximately $M_Z/2$ and its variation
is very effectively reduced.  The shape of the cross section variation at the 
LHC on the other hand is qualitatively unchanged when going from LO
to NLO.  We attribute this to new channels that become active at NLO 
(see Sec.~\ref{real-subsection}), which have a modest impact 
at the Tevatron, but a sizable impact at the LHC, due to parton 
densities being probed in different $x$ regions.  One might be tempted
to conclude that a constant $K$-factor is a good approximation for
the inclusive NLO cross section at the LHC.  However, the $K$ factor 
(also shown) varies between 1.3 and 1.6 in the displayed scale range.
Following Ref.~\cite{Dittmaier:2007th}, we also calculate an exclusive
NLO cross section for the LHC by vetoing 2-jet events with a second hardest jet 
with $p_T > 50$ GeV (NLO with 2$^\text{\scriptsize nd}$ jet veto).
This exclusive NLO LHC cross section decreases for scales below $M_Z$
and has a strongly reduced scale uncertainty.  Since the qualitative
features of our results are similar to those found for $WWj$ production,
we quantify the $ZZj$ LO and NLO scale uncertainties for the Tevatron
and LHC in Tables~\ref{tbl:scaleuncertainty_tevatron} and 
\ref{tbl:scaleuncertainty_lhc}, respectively.
\begin{table}[tb]
\centerline{
\def\arraystretch{1.2}
\begin{tabular}{|c|c|c|c|}
\cline{2-4}
\multicolumn{1}{c}{} & \multicolumn{3}{|c|}{$\Delta\sigma/\sigma(p\bar{p}\to ZZ+\text{jet})$, $\sqrt{s}=1.96$ TeV} \\
\cline{2-4}
\multicolumn{1}{c|}{} & \hspace*{0.2cm}$\mu/M_Z\in[\frac{1}{2}, 2]$\hspace*{0.2cm} & \hspace*{0.2cm}$\mu/M_Z\in[\frac{1}{4}, 4]$\hspace*{0.2cm} & \hspace*{0.2cm}$\mu/M_Z\in[\frac{1}{8}, 8]$\hspace*{0.2cm} \\
\hline
LO  & $23\%$ & $44\%$ & $62\%$ \\
\hspace*{0.2cm}NLO\hspace*{0.2cm} & $6\%$ & $11\%$ & $19\%$ \\
\hline
\end{tabular}}
\vspace*{.5cm}
\caption{\label{tbl:scaleuncertainty_tevatron}
Scale uncertainty for LO and NLO cross sections for $ZZ+\text{jet}$ production 
at the Tevatron as function of the scale variation.
The relative scale uncertainty is defined through the envelope:
$\Delta\sigma/\sigma:=[\sigma_\text{max}(\mu\in I)-\sigma_\text{min}(\mu\in I)]/[\sigma_\text{max}(\mu\in I)+\sigma_\text{min}(\mu\in I)]$.
Other details as in Fig.~\ref{fig:scalevar}.}
\vspace*{0.3cm}
\end{table}
\begin{table}[tb]
\centerline{
\def\arraystretch{1.2}
\begin{tabular}{|c|c|c|c|}
\cline{2-4}
\multicolumn{1}{c}{} & \multicolumn{3}{|c|}{$\Delta\sigma/\sigma(pp\to ZZ+\text{jet})$, $\sqrt{s}=14$ TeV} \\
\cline{2-4}
\multicolumn{1}{c|}{} & \hspace*{0.2cm}$\mu/M_Z\in[\frac{1}{2}, 2]$\hspace*{0.2cm} & \hspace*{0.2cm}$\mu/M_Z\in[\frac{1}{4}, 4]$\hspace*{0.2cm} & \hspace*{0.2cm}$\mu/M_Z\in[\frac{1}{8}, 8]$\hspace*{0.2cm} \\
\hline
LO & $12\%$ & $23\%$ & $34\%$ \\
NLO & $7\%$ & $15\%$ & $23\%$ \\
\hspace*{0.2cm}NLO with $2^\text{nd}$ jet veto\hspace*{0.2cm} & $0.5\%$ & $3\%$ & $6\%$ \\
\hline
\end{tabular}}
\vspace*{.5cm}
\caption{\label{tbl:scaleuncertainty_lhc}
Scale uncertainty for LO and NLO cross sections for $ZZ+\text{jet}$ production 
at the LHC as function of the scale variation.
Other details as in Fig.~\ref{fig:scalevar} and Table \ref{tbl:scaleuncertainty_tevatron}.
}
\end{table}
When comparing our results 
for $\mu/M_Z\in[\frac{1}{2}, 2]$ with the corresponding $WWj$ results extracted 
from Tables 4 and 1 in Ref.~\cite{Dittmaier:2009un}, we find deviations of 
less than 2 percentage points.

\begin{figure}
\begin{center}
\hspace*{0.cm}\begin{minipage}[c]{.49\linewidth}
\flushleft \includegraphics[height=9.5cm,angle=0,clip=true]{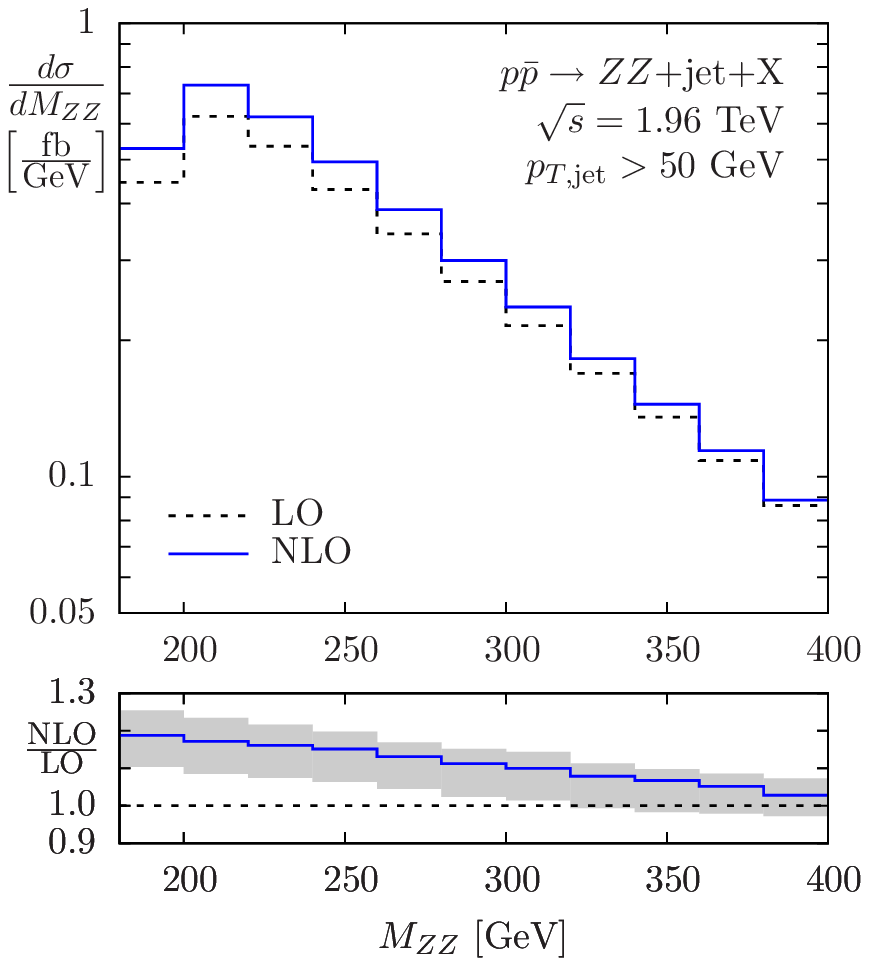}
\end{minipage}\hspace{0.05cm}
\begin{minipage}[c]{.49\linewidth}
\flushright \includegraphics[height=9.5cm,angle=0,clip=true]{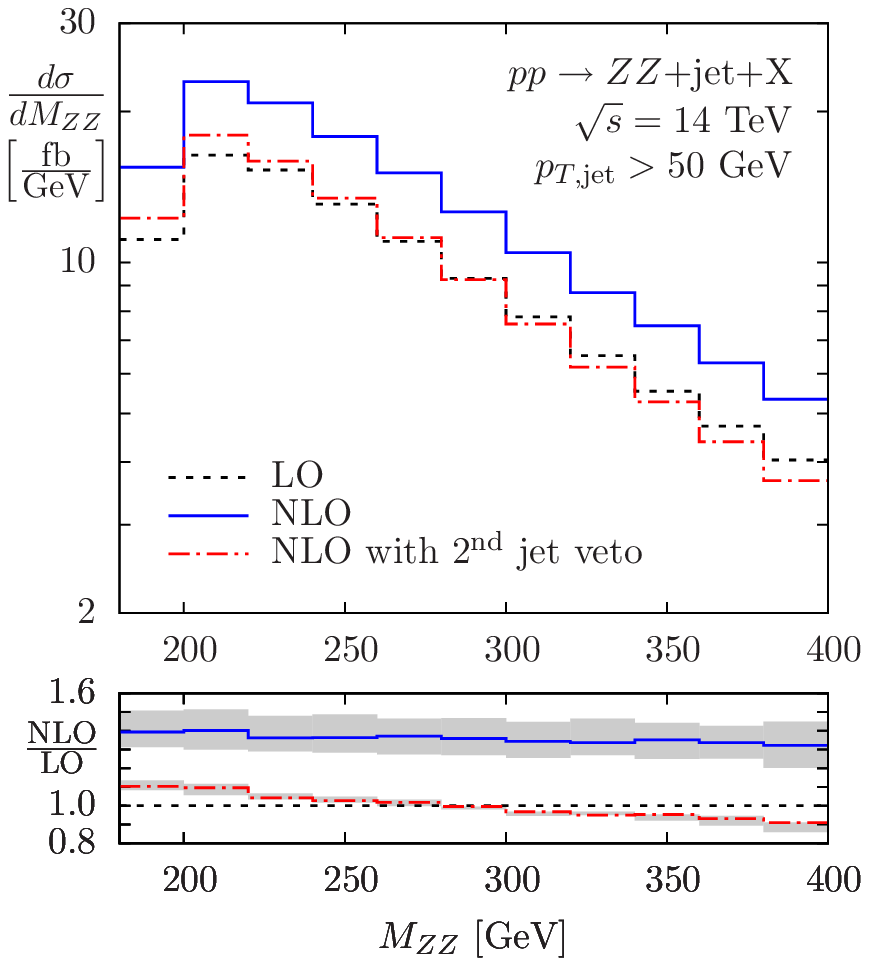}
\end{minipage}\\[0.2cm]
\caption{%
$ZZ$ invariant mass distribution for $ZZ$+jet production at the Tevatron and 
LHC with $\mu_R=\mu_F=M_Z$.  The differential $K$ factor
is also shown.  The $K$-factor bands are defined in the main text.  
Other details as in Fig.~\protect\ref{fig:scalevar}.
\label{fig:mzz}}
\end{center}
\end{figure}

In Tables \ref{tbl:ptcut_tevatron} and \ref{tbl:ptcut_lhc}, we show 
for the Tevatron and LHC, respectively, how the
$ZZj$ LO and NLO cross sections change when the $p_T$ cut on the hardest jet
is varied.
\begin{table}[tb]
\centerline{
\def\arraystretch{1.2}
\begin{tabular}{|c|c|c|c|c|}
\cline{2-5}
\multicolumn{1}{c}{} & \multicolumn{4}{|c|}{$\sigma(p\bar{p}\to ZZ+\text{jet})$ [pb], $\sqrt{s}=1.96$ TeV} \\
\hline
\hspace*{0.2cm}$p_{T,\text{jet}}$ cut [GeV]\hspace*{0.2cm} & $20$ & $50$ & $100$ & $200$ \\
\hline
LO & $0.27202(3)$ & $0.07456(1)^{+28\%}_{-20\%}$ &  $0.016037(2)$ & $0.0012651(1)$ \\
NLO & $0.3307(6)$ & $0.0836(1)^{+5\%}_{-7\%}$ &  $0.01583(4)$ &  $0.000976(4)$ \\
\hline
\end{tabular}}
\vspace*{.5cm}
\caption{\label{tbl:ptcut_tevatron}
$ZZ$+jet production cross section at the Tevatron with different $p_T$ cuts 
for the hardest jet.  The scale $\mu=M_Z$ is employed with $\mu_R=\mu_F=\mu$. 
The integration error is given in brackets.  The minimum and maximum relative deviation from $\sigma(\mu=M_Z)$ obtained with the scale variation $\mu/M_Z\in[\frac{1}{2}, 2]$ is shown as sub- and superscript for a $p_{T,\text{jet}}$ cut of $50$ GeV.
Input parameters are defined in the main text.}
\end{table}
\begin{table}[tb]
\centerline{
\def\arraystretch{1.2}
\begin{tabular}{|c|c|c|c|c|}
\cline{2-5}
\multicolumn{1}{c}{} & \multicolumn{4}{|c|}{\hspace*{0.2cm}$\sigma(pp\to ZZ+\text{jet})$ [pb], $\sqrt{s}=14$ TeV\hspace*{0.2cm}} \\
\hline
$p_{T,\text{jet}}$ cut [GeV] & $20$ & $50$ & $100$ & $200$ \\
\hline
LO & $6.505(1)$ & $2.6978(4)^{+13\%}_{-11\%}$ &  $1.0066(1)$ & $0.22974(3)$ \\
NLO & $8.01(3)$ & $3.653(9)^{+8\%}_{-6\%}$ &  $1.511(4)$ &  $0.415(2)$ \\
\hspace*{0.2cm}NLO with $2^\text{nd}$ jet veto\hspace*{0.2cm} & & $2.637(9)^{+0.2\%}_{-1\%}$ &  $0.755(4)$ &  $0.1005(9)$ \\
\hline
\end{tabular}}
\vspace*{.5cm}
\caption{\label{tbl:ptcut_lhc}
$ZZ$+jet production cross section at the LHC with different $p_T$ cuts 
for the hardest jet.  For cut values above 20 GeV, we also give 
the NLO cross section when a $p_{T,\text{jet}} > 50$ GeV veto for 
additional jets is added to the selection.
Other details as in Table \ref{tbl:ptcut_tevatron}.}
\end{table}
These results demonstrate that all $K$ factors are $p_T$ cut dependent.
In general, the $K$ factor for $ZZj$ production will have a non-negligible 
differential dependence.  As example, we display in Fig.~\ref{fig:mzz} the 
differential LO and NLO distributions with respect to the invariant $ZZ$ mass 
and the resulting $K$ factor at the Tevatron and LHC.
The $K$-factor bands shown in this figure correspond to a variation of
the scale $\mu$ by a factor of $2$ in the NLO differential cross section only,
i.e.~we display $[d\sigma_\text{NLO}/dM_{ZZ}](\mu)/[d\sigma_\text{LO}/dM_{ZZ}](M_Z)$
with $\mu/M_Z\in[\frac{1}{2}, 2]$.  One can distinguish the modest variation of the 
inclusive NLO $K$ factor at the LHC from the strong decrease of the other $K$ factors 
with increasing $M_{ZZ}$.


\section{Conclusions}
In this paper we have presented first results for $ZZj$ production at NLO QCD
accuracy, obtained with fully differential parton-level Monte Carlo programs that 
allow to 
take into account realistic experimental selection cuts.  For a default
scale choice of $\mu = M_Z$ and a $p_T$ cut of 50 GeV for the hardest jet
we find a $K$ factor of 1.1 and 1.35 at the Tevatron and LHC, respectively.
At the Tevatron, the NLO corrections stabilise the LO prediction for 
cross sections considerably.  The shape of the cross section variation at the 
LHC on the other hand is qualitatively unchanged when going from LO
to NLO.  Nevertheless, at the LHC stabilisation at NLO can still be achieved by  
applying suitable selection cuts like for example a veto against the emission of a 
second hard jet.  Our results indicate that residual scale uncertainties are 
typically at the 10\% level and can be further reduced to about 5\% or less 
by applying suitable selections.


\begin{acknowledgments}
We would like to thank N.~Greiner, J.P.~Guillet, P.~Uwer and M.~Worek
for helpful discussions and contributions.  
T.B.~is supported in parts by the Science and Technology Facilities Council
(STFC) and the Scottish Universities Physics Alliance (SUPA).
T.G.'s work is supported by the US Department of Energy, contract DE-AC02-76SF00515.
N.K.~thanks the Higher Education Funding Council for England and STFC for 
financial support under the SEPnet Initiative.
IPPP Associates T.B.~and N.K.~thank the Institute for Particle Physics 
Phenomenology (IPPP) Durham for support.
This work has been supported in part by the the DFG Graduiertenkolleg ``Elementary Particle Physics at the TeV Scale'', the Helmholtz Alliance ``Physics at the Terascale'' and the European Community's Marie-Curie Research Training Network under contract MRTNCT-2006-035505 ``Tools and Precision Calculations for Physics Discoveries at Colliders''. 
T.B.~and N.K.~thank the Galileo Galilei Institute for Theoretical
Physics for the hospitality and the INFN for partial support during the
completion of this work.
This work has made use of the resources provided by the Edinburgh 
Compute and Data Facility (ECDF).  The ECDF is partially 
supported by the eDIKT Initiative.
This work was supported by the BMBF and DFG, Germany, contracts 05HT1WWA2 and BI 1050/2.
Jaxodraw \cite{Jaxodraw} was used to draw the Feynman diagrams in this paper.
\end{acknowledgments}

\end{document}